# Natural Selection between Two Games and Replicator Dynamics on Graphs

G. Karev

National Center for Biotechnology Information, National Institutes of Health, Bethesda, MD 20894, USA (retired); gkarev@gmail.com

**Introduction**

Game theory can provide insights into the outcome of interactions between individuals or groups of individuals, if the rules for how they make decisions as a result of their interactions, given their own background information, are understood. Typically, it is assumed that individuals will act in a way that will maximize their "payoff", however that may be defined in a specific situation. In evolutionary game theory, the payoff to be maximized is "fitness", i.e., the ability to propagate and maximize the number of offspring. If a characteristic that determines the size of the "payoff" is heritable, then it is assumed that it will be passed along to be used by the offspring in the next iteration.

It is typically assumed that interacting individuals are playing the same game, i.e., will make payoff decisions based on the same underlying set of assumptions. Classic examples of such games are Prisoner's Dilemma (PD), Harmony (H), Hawk-Dove (HD), also known as a game of chicken, or Stag Hunt (SH) also known as coordination game. Depending on the specifics of the game, the individuals will make different decisions on whether to cooperate or defect in order to maximize their payoff, with pure strategies (only cooperation of defection) dominating in PD or H games, mixed strategies dominating in HD, and any of pure strategies in SH game.

The question of what will emerge as the dominant strategy in a situation, where some individuals are playing one game, and others are playing another, has been partially explored in, i.e., (Kareva, Karev, 2019) in the context of natural selection between game theoretical models of cancer. Here we will address the question of which game, rather than which strategy, will dominate over time as a result of natural selection within a community, where different individuals can play different games at the same time.

A mathematical "toolbox", that allows us to study this process was suggested in (Kareva, Karev, 2019) and developed further by (Karev, 2024) where the approach was applied to some examples from the literature, including local replicator dynamics (C. Hilbe, 2011), pairwise competition (J. Morgan, K. Steglitz, 2003) and the game of alleles in diploid genomes (Bohl et.al, 2014).

Here I apply this approach to evolutionary games on graphs. In these games, individuals occupy the vertices of a graph and can obtain a payoff from interaction with all adjacent neighbors. In addition to the traditional case of a "complete graph", where all vertices are connected to each other, Ohtsuki and Nowak (2006) considered three different update rules for the evolutionary dynamics on graphs (see s.3 below). In this paper I explore the question of which game - between the classic and the Ohtsuki-Nowak updated versions – is "better", i.e. will come to dominate over time.

1. **Replicator equations and parametrized games**

Assume the individuals in a population can choose among $n$ strategies, and define $l_i(t)$ as the number of individuals ($i$-players) playing $i$-th strategy at time $t$. An $i$-player interacting with a $j$-player obtains payoff $a_{ij}$ ; $A = (a_{ij}), i, j = 1 \ldots n$ is then defined as the payoff matrix for all pairwise interactions between the players. Let us denote $N(t) = \sum_{i=1}^{n} l_i$ the total population size and $x_i = l_i/N$ the frequency of $i$-players. If the population is infinite and well-mixed, then the frequencies of $i$-players solve the replicator equation

$$\frac{dx_i}{dt} = x_i[(Ax)_i - xAx]. \tag{1.1}$$

Here $(Ax)_i = \sum_{j=1}^{n} a_{ij} x_j$ is the expected fitness of strategy $i$ (or the payoff obtained by the $i$-th strategist after interaction with the total population) and $xAx = \sum_{i,j=1}^{n} x_i a_{ij} x_j$ is the average payoff over the entire population (see Taylor and Jonker 1978; Hofbauer and Sigmund 1998 for details).

In what follows, we consider two different 2-player games, which we refer to as $g0$ and $g1$, with corresponding payoff matrices $G^0$ and $G^1$. Let us define the "$\beta$-population" composed of individuals who can play the $g1$-game with a certain hereditary probability $\beta$ and the $g0$-game with the hereditary probability $(1-\beta)$. Individuals in the $\beta$-population play the "mixed" game with the parametrized payoff matrix

$$G^\beta = (1-\beta)G^0 + \beta G^1. \tag{1.2}$$

Let us consider a community composed of all $\beta$-populations, where $0 \leq \beta \leq 1$. Our objective is to determine which game, $g1$, $g0$, or a mixture thereof, will dominate over time as a result of natural selection within this community. Evolution of the distribution of the parameter $\beta$ can provide us the answer to this question.

## 2. Mathematical Toolbox

In what follows, we consider 2-player games with 2 possible strategies, $s_0$ and $s_1$.

Assume

$$G^0 = \begin{bmatrix} a & b \\ c & d \end{bmatrix}, \quad G^1 = \begin{bmatrix} e & f \\ g & h \end{bmatrix}.$$

Then the payoff matrix for the mixed game is

$$G^\beta = \begin{bmatrix} G_{00}^{(\beta)} & G_{01}^{(\beta)} \\ G_{10}^{(\beta)} & G_{11}^{(\beta)} \end{bmatrix} = (1-\beta)G^0 + \beta G^1 = \begin{bmatrix} a(1-\beta)+\beta e & b(1-\beta)+\beta f \\ c(1-\beta)+\beta g & d(1-\beta)+\beta h \end{bmatrix}. \quad (2.1)$$

The question of interest is the evolution of the distribution of parameter $\beta$ over time.

Let us assume, for simplicity, that players in all populations adopt the $s_1$ strategy with probability $p_1$ and adopt the $s_0$ strategy with probability $p_0$ at the initial time $t = 0$.

Let us now formally introduce the auxiliary keystone variable $q(t)$ using the following equation:

$$\frac{dq(t)}{dt} = x(t), q(0) = 0.$$

Explicit form of this equation is given by equation (2.3) below.

Now let us denote

$$W_0(t, \beta) = (-G_{00}^{(\beta)} + G_{01}^{(\beta)}) q(t) + G_{00}^{(\beta)} t,$$

$$W_1(t, \beta) = (-G_{10}^{(\beta)} + G_{11}^{(\beta)}) q(t) + G_{10}^{(\beta)} t,$$

$$V_0(t) = \int_B \exp(W_0(t, \beta)) P(0, \beta) d\beta,$$

$$V_1(t) = \int_B \exp(W_1(t, \beta)) P(0; \beta) d\beta.$$

Here $P(0; \beta)$ is the distribution of parameter $\beta$ at the initial time.

Then the pdf of the parameter $\beta$ at $t$ time is

$$P(t, \beta) = P(0, \beta) \left( \frac{p_0 \exp(W_0(t,\beta)) + p_1 \exp(W_1(t,\beta))}{p_0 V_0(t) + p_1 V_1(t)} \right) \quad (2.2)$$

and the variable $q(t)$ solves the explicit equation

$$\frac{dq}{dt} = \frac{p_1 V_1(t)}{p_0 V_0(t) + p_1 V_1(t)}. \tag{2.3}$$

See (Karev, 2024) for proof and some applications.

## 3. Games on graphs. Natural selection between initial and updated games

A possible application of the developed approach is to evolutionary games on graphs. In these games, individuals occupy the vertices of a graph and receive payoffs from interaction with all adjacent neighbors. The traditional replicator equation (1.1) describes the case of a "complete graph", where all vertices are connected to each other.

Ohtsuki and Nowak (2006) considered three different update rules for the evolutionary dynamics on graphs of degree *k*, which they call 'birth-death' (BD), 'death-birth' (DB) and 'imitation' (IM).

(i) For BD updating, an individual is selected for reproduction from the entire population proportional to fitness; the offspring of this individual replaces a randomly chosen neighbor.

(ii) For DB updating, a random individual from the entire population is chosen to die; the neighbors compete for the b empty site proportional to fitness.

(iii) For IM updating, a random individual from the entire population is chosen to revise its strategy; it will either keep its current strategy or imitate one of the neighbors' strategies proportional to fitness.

In what follows we refer to the games that use the corresponding rules as BD, DB, and IM games, respectively. We will then evaluate the interactions between the games with update rules, and the original game.

Let $A = (a_{ij}), i, j = 1 \ldots n$ be the payoff matrix of some underlying game, and let k be the degree of the graph. Then the three updated mechanisms BD, DB, and IM correspond to matrices

$$BD = \left(\frac{a_{ii} + a_{ij} - a_{ji} - a_{jj}}{k-2}\right), \tag{3.1}$$

$$DB = \left(\frac{(k+1)a_{ii} + a_{ij} - a_{ji} - (k+1)a_{jj}}{(k+1)(k-2)}\right), \tag{3.2}$$

$$IN = \left(\frac{(k+3)a_{ii}+3a_{ij}-3a_{ji}-(k+3)a_{jj}}{(k+3)(k-2)}\right). \tag{3.3}$$

Let $x_i(t)$ be the expected frequency of strategy $i$ on an infinitely large graph of degree $k$ at time $t$. Ohtsuki and Nowak (2006) proved that $x_i(t)$ solves the 'replicator equations on graphs', which coincide with the standard replicator equation with a modified payoff matrices $A + BD$, $A + DB$ or $A + IM$ corresponding to 'birth-death' (BD), 'death-birth' (DB) and 'imitation' (IM) games on graphs respectively.

To answer the question of what game between original game (no updates) and game with updates, will dominate, let us consider a model of a community composed of $\beta$- populations with different payoff matrices $G^\beta$ of the form (2.1) and study the natural selection between the populations. This process can be formally described by the evolution of the distribution of parameter $\beta$ and the outcome of the evolution determines the "dominant" game. The toolbox for studying this problem was summarized above.

Here we consider the games with only 2 strategies and the payoff matrix $A = (a_{ij}), i,j = 1,2$. In this case,

$$BD = \frac{1}{k-2}\begin{bmatrix} 0 & B \\ -B & 0 \end{bmatrix}, \quad \text{where} \quad B = a_{11} + a_{12} - a_{21} - a_{22}; \tag{3.4}$$

$$DB = \frac{1}{(k+1)(k-2)}\begin{bmatrix} 0 & D \\ -D & 0 \end{bmatrix}, \text{ where } D = (k+1)a_{11} + a_{12} - a_{21} - (k+1)a_{22}; \tag{3.5}$$

$$IN = \frac{1}{(k+3)(k-2)}\begin{bmatrix} 0 & I \\ -I & 0 \end{bmatrix}, \text{ where } I = (k+3)a_{11} + 3a_{12} - 3a_{21} - (k+3)a_{22}. \tag{3.6}$$

Below we consider the natural selection between initial and updated games on graphs with $k=3$.

**Example 3.1.**

Let the initial game be the Prisoner's Dilemma with the payoff matrix:

$$A = \begin{bmatrix} 5 & 0 \\ 8 & 1 \end{bmatrix} \tag{3.7}$$

(see Outsuki, Nowak, Eq. (38)). Then corresponding update matrices (3.5)-(3.7) are

$$BD = \begin{bmatrix} 0 & -4 \\ 4 & 0 \end{bmatrix}, \tag{3.8}$$

$$DB = \begin{bmatrix} 0 & 2 \\ -2 & 0 \end{bmatrix}, \tag{3.9}$$

$$IN = \begin{bmatrix} 0 & 0 \\ 0 & 0 \end{bmatrix}. \tag{3.10}$$

Let us consider selection between the original game $A$ and the $BD$ game. Corresponding payoff matrix for the mixed game is

$$G^\beta = (1 - \beta)A + \beta(A + BD) = A + \beta BD = \begin{bmatrix} 5 & -4\beta \\ 8 + 4\beta & 1 \end{bmatrix}. \tag{3.11}$$

The selection process is described by the dynamics of the distribution of parameter $\beta$ for parametric game (3.11), see Fig.4..

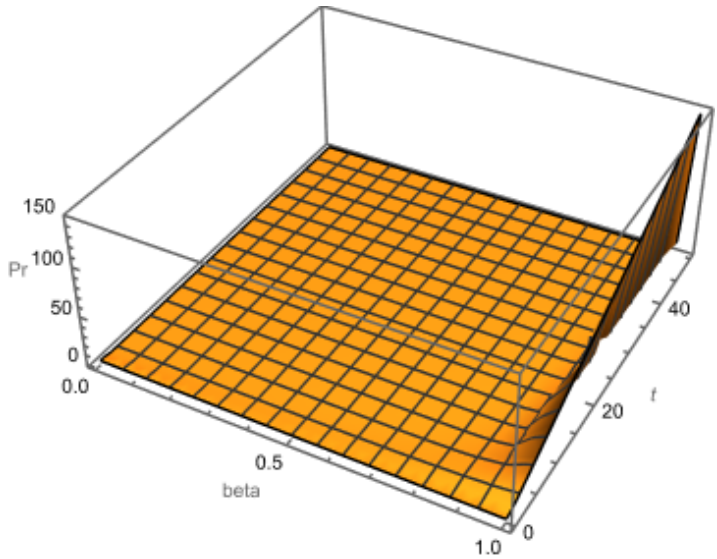

Fig. 4. Dynamics of the distribution of parameter $\beta$ for parametric game (3.11).

The distribution tends to concentrate at point $\beta = 1$ indicating that the BD game dominates over the original game.

Next, let us consider selection between the original game (3.7) and the $DB$ game (3.9). The corresponding payoff matrix for the mixed game is

$$G^\beta = (1 - \beta)A + \beta(A + DB) = A + \beta DB = \begin{bmatrix} 5 & 2\beta \\ 8 - 2\beta & 1 \end{bmatrix}. \tag{3.12}$$

Dynamics of the distribution of parameter $\beta$ is shown on the Fig.5.

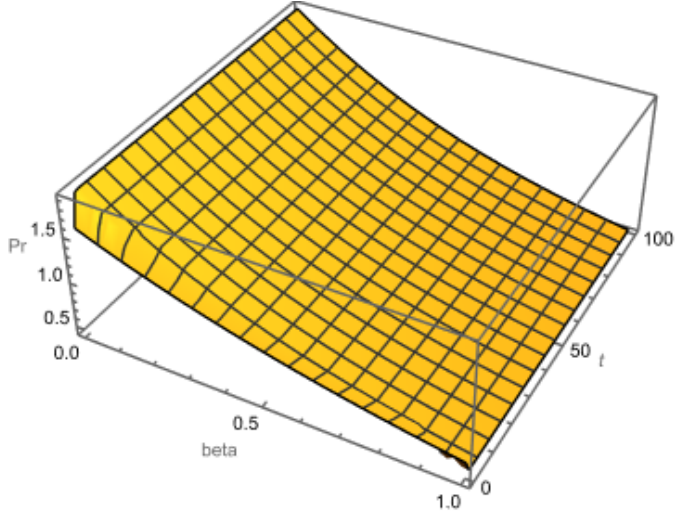

Fig.5. Selection between the original game *A* (3.7) and *DB* game (3.9): dynamics of the distribution of parameter $\beta$ for parametric game (3.12).

We can see that the distribution of the parameter $\beta$ is non-degenerated at all time; this means that individuals who play mixed games *A* and *DB* (corresponding to the parameter values $0 < \beta < 1$) persist in the community indefinitely.

**Example 3.2.**

Now assume the payoff matrix for the original game is given by the following (see Outsuki, Nowak, Eq. (40)):

$$A = \begin{bmatrix} 15 & 0 \\ 16 & 8 \end{bmatrix}. \tag{3.13}$$

This game is also Prisoner's Dilemma. The corresponding matrices for the updated games (3.4)-(3.6) are

$$BD = \begin{bmatrix} 0 & -9 \\ 9 & 0 \end{bmatrix}, \tag{3.14}$$

$$DB = \begin{bmatrix} 0 & 12 \\ -12 & 0 \end{bmatrix}, \tag{3.15}$$

$$IN = \begin{bmatrix} 0 & -6 \\ 6 & 0 \end{bmatrix}. \tag{3.16}$$

The selection dynamics between original game (3.13) and corresponding BD game (3.14) are defined by the parametrized payoff matrix

$$G^\beta = (1-\beta)A + \beta(A+BD) = A + \beta BD = \begin{bmatrix} 15 & -9\beta \\ 16+9\beta & 8 \end{bmatrix} \quad (3.17)$$

The dynamics of the distribution of parameter $\beta$ for parametrized game (3.17) is shown in the Fig.6. Over time, the distribution tends to concentrate at the point $\beta = 1$, indicated that natural selection favors the $BD$ game over the original game.

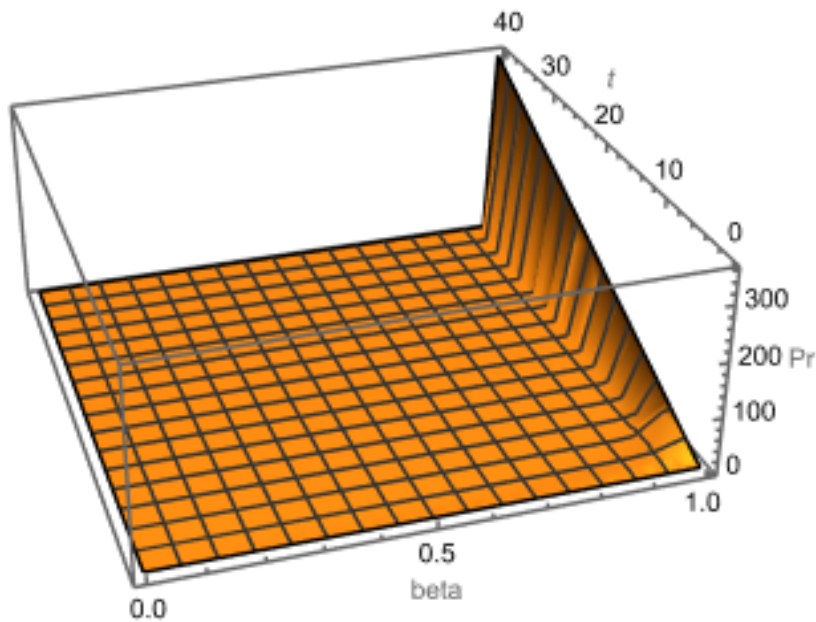

Fig.6. Selection between the original game $A$ (3.13) and the $BD$ game (3.14).

The parametrized game for selection between the original game (3.13) and corresponding $DB$ game (3.15) is defined by the payoff matrix

$$G^\beta = (1-\beta)A + \beta(A+DB) = A + \beta DB = \begin{bmatrix} 15 & 12\beta \\ 16-12\beta & 8 \end{bmatrix} \quad (3.18)$$

Dynamics of the distribution of parameter $\beta$ for parametrized game (3.18) are shown in Figure 7.

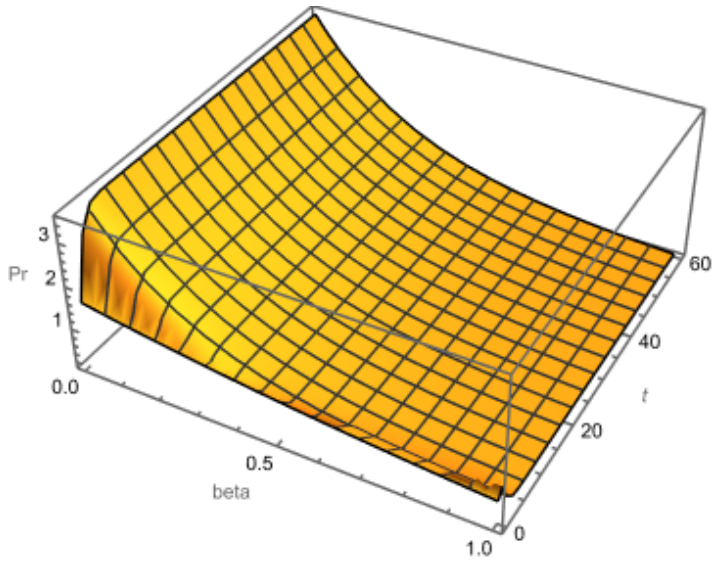

Fig. 7. Selection between the original game *A* (3.13) and *DB* game (3.15): dynamics of the parameter $\beta$ distribution for parametric game (3.18).

The distribution of parameter $\beta$ remains non-degenerated at all times indicated that individuals who play mixed games A and DB (corresponding to parameter values $0 < \beta < 1$) persist in the community indefinitely.

Finally, the parametrized game for the selection between the original game (3.13) and the corresponding *IN* game (3.16) is determined by the payoff matrix

$$G^\beta = A + \beta IN = \begin{bmatrix} 15 & -6\beta \\ 16 + 6\beta & 8 \end{bmatrix}. \tag{3.19}$$

Dynamics of the distribution of parameter $\beta$ in parametrized game (3.19) are shown in Fig.8.

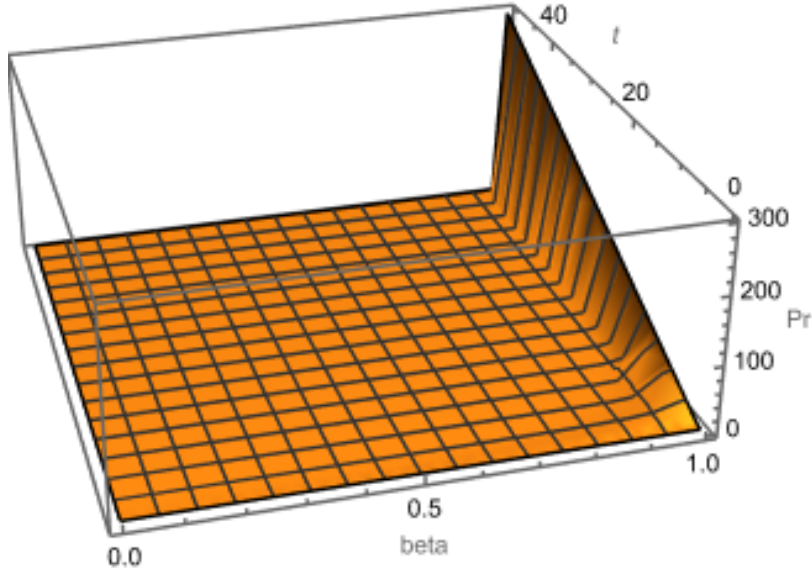

Fig.8. Selection between original game *A* (3.13) and *IN* game (3.16): dynamics of the distribution of parameter $\beta$ for parametric game (3.19).

One can see that over time, the distribution tends to concentrate at the point $\beta = 1$, indicated that natural selection favors the *IN* game over the original game.

To summarize, in these examples we looked at which evolutionary game on a graph would dominate over time. We considered the cases with update rules described by Ohtsuki and Nowak (2006), called BD (birth-death), DB (death-birth) and IM (imitation), and evaluated whether a game with update rules would dominate over a game without update rules (the original game A) as measured by evolution of parameter $\beta$ in the community.

## 4. Games on graphs. Natural selection between games with different updating rules

Now let us consider the selection between games with different update rules. The "mix" of *BD* and *DB* games (let's call it BDDB) corresponds to the parametrized payoff matrix

$$G^\beta = (A + BD)(1 - \beta) + \beta (A + DB) = A + BD (1 - \beta) + \beta\, DB, \qquad (4.1)$$

$0 \leq \beta \leq 1$.

The mix of *BD* and *IN* games (let's call it BDIN) corresponds to the parametrized payoff matrix

$$G^\beta = A + BD (1 - \beta) + \beta\, IN \qquad (4.2)$$

The mix of *DB* and *IN* games (let's call it DBIN) correspond to the parametrized payoff matrix

$$G^\beta = A + DB\,(1-\beta) + \beta\,IN. \qquad (4.3)$$

In what follows we study the evolution of the distribution of parameter $\beta$ for parametrized games (4.1)-(4.3).

**Example 4.1.**

Let us consider the natural selection between BD and DB games with payoff matrixes (3.8) and (3.9), given an initial game with the payoff matrix $A = \begin{bmatrix} 5 & 0 \\ 8 & 1 \end{bmatrix}$.

In this case

$$G^\beta = A + (1-\beta)\,BD + \beta\,DB = \begin{bmatrix} 5 & -4+6\beta \\ 12-6\beta & 1 \end{bmatrix}. \qquad (4.4)$$

The following Figure 9 shows the evolution of the distribution of parameter $\beta$ for the parametrized BDDB game (4.4).

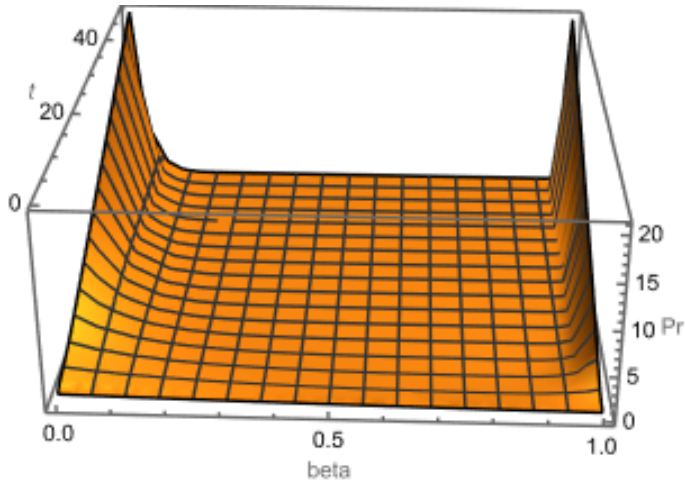

Fig.9. Evolution of the distribution of parameter $\beta$ for the parametrized BDDB game (4.4).

We can see that, in this example, all individuals who play mixed games (corresponding to the parameter values $0 < \beta < 1$) are eliminated from the community over time. However, individuals playing pure games (*BD* corresponding to the parameter value $\beta = 0$ and *DB* corresponding to $\beta = 1$) persist in the community indefinitely, maintaining some proportion $R(t) = P(t,1)/P(t,0)$.

The plot of $R(t)$ for the parametrized BDDB game (4.4) is shown in Fig.10.

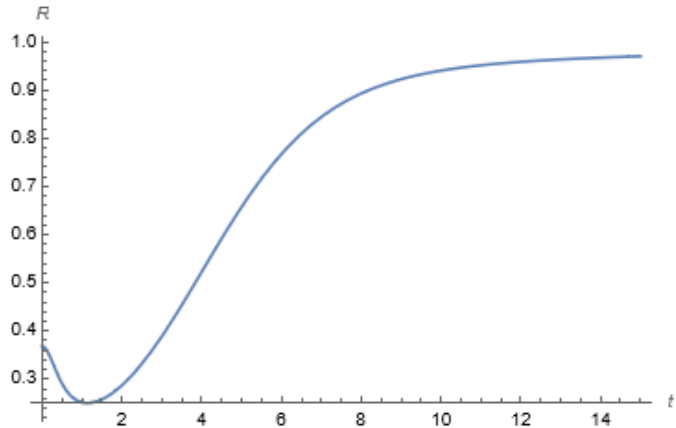

Fig.10. Proportion of the individuals playing pure games $R(t) = P(t,1)/P(t,0)$ for the parametrized BDDB game (4.4).

Now, let us look at the BDIN parametrized mix game, which describes natural selection between BD and IN games given by matrixes (3.8) and (3.10). The outcome of its evolution over time is determined by the payoff matrix

$$G^\beta = A + BD\,(1-\beta) + \beta\,IN = \begin{bmatrix} 5 & -4(1-\beta) \\ 12 - 4\beta & 1 \end{bmatrix}. \quad (4.5)$$

The following Figure 11 shows the evolution of the distribution of parameter $\beta$ for the BDIN parametrized game (4.5).

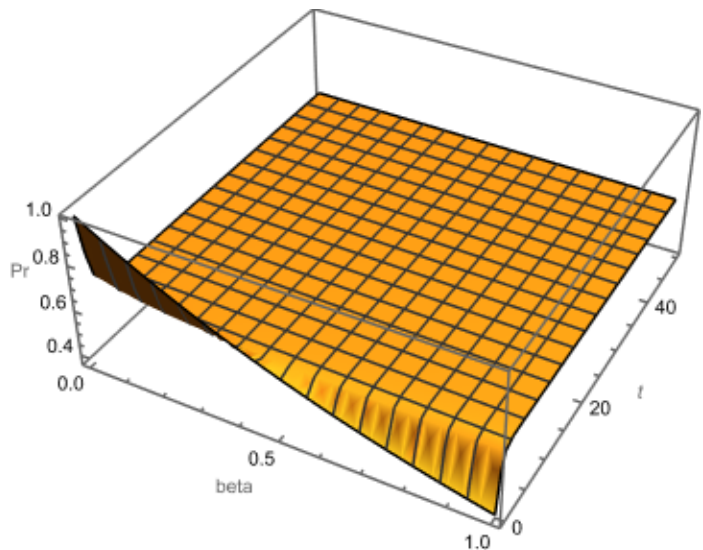

Fig.11. Evolution of the distribution pf parameter of $\beta$ for BDIN parametrized game (4.5).

We can see that the distribution of parameter $\beta$ is non-degenerated at all times. This means that individuals who play a mix of *BD* and *IN* games (corresponding to the parameter values $0 < \beta < 1$) persist in the community indefinitely.

Finally, for the mixed DBIN game the parametrized payoff matrix

$$G^\beta = A + DB(1-\beta) + \beta IN = \begin{bmatrix} 5 & 2(1-\beta) \\ 6+2\beta & 1 \end{bmatrix}. \tag{4.6}$$

The following Figure 12 shows the evolution of the distribution of parameter $\beta$ for the DBIN parametrized game (4.6).

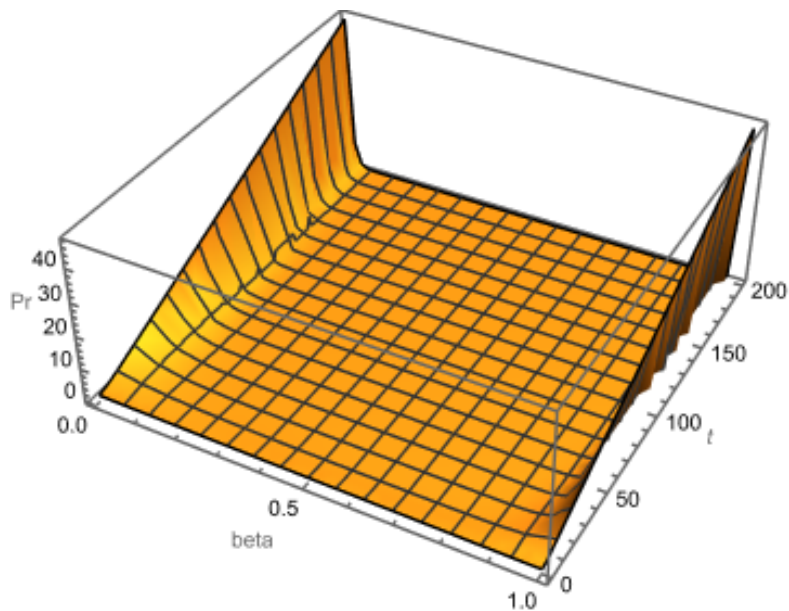

Figure 12. Evolution of the distribution of parameter $\beta$ for the DBIN parametrized game (4.6).

Similar to the parametrized BDB game (4.4), all individuals who play mixed games (corresponding to the parameter values $0 < \beta < 1$) are eliminated from the community over time; however, the individuals playing pure games (*DB* corresponding to the parameter value $\beta = 0$ and *IN* corresponding to $\beta = 1$) persist in the community indefinitely. The plot of the proportion of individuals playing pure games $R(t) = P(t,1)/P(t,0)$ for the parametrized DBIN game (4.6) is shown in Fig.13.

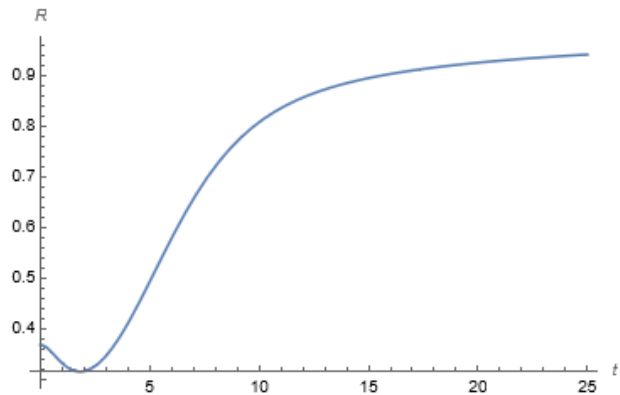

Fig.13. Proportion of the individuals playing pure games $R(t) = P(t, 1)/P(t, 0)$ for the DBIN parametrized game (4.6).

**Conclusion**

In this paper I study the natural selection between two games to determine, which game will dominate in the community as a result of natural selection. The formalization of this question in the form of a "parametrized game" and the mathematical toolbox to solve it were recently developed in (Karev 2024). Here, I applied the developed approach to natural selection between games on graphs. Specifically, I compare standard games on a complete graph with games that have updated rules suggested by Ohtsuki and Nowak (2006). The results of the natural selection between the initial and updated games, or between different updated games were visualized through the evolution of the parameter distribution in corresponding parametrized game. The developed approach can be used for other problems, such as local replicator dynamics, pairwise competition, the game of alleles in diploid genomes (see Karev 2004) and cancer modeling (Kareva, Karev 2019).

**References**
K. Bohl, S. Hummert, S. Werner, D. Basanta, A. Deutsch, S. Schuster, G. Theißen and A. Schroeter (2014). Evolutionary game theory: molecules as players. Mol. BioSyst, 10, 3066.